\documentclass[reprint,superscriptaddress,amsmath,amssymb,aps]{revtex4-2}
\usepackage{physics}
\usepackage{graphicx} 
\usepackage{dcolumn}  
\usepackage{bm}       
\usepackage{hyperref} 
\usepackage{romannum}
\usepackage{xcolor}

\begin{document}
\pagenumbering{arabic}
\newcommand{\angstrom}{\mbox{\normalfont\AA}}
\renewcommand{\thetable}{\Roman{table}}
\title{Robustness of the intrinsic anomalous Hall effect in \texorpdfstring{$\mathrm{Fe}_{3}\mathrm{GeTe}_{2}$}{Lg} to a uniaxial strain}

\author{Mijin Lim}
 \thanks{These authors contributed equally to this work}
 \affiliation{Department of Physics, Pohang University of Science and Technology, Pohang 37673, Korea}
\author{Byeonghyeon Choi}
 \thanks{These authors contributed equally to this work}
 \affiliation{Department of Physics, Pohang University of Science and Technology, Pohang 37673, Korea}
\author{Minjae Ghim}
 \affiliation{Department of Physics, Pohang University of Science and Technology, Pohang 37673, Korea}
\author{Je-Geun Park}
 \affiliation{Department of Physics and Astronomy, Seoul National University, Seoul 08826, Korea}
\author{Hyun-Woo Lee}
 \email{Corresponding author: hwl@postech.ac.kr}
 \affiliation{Department of Physics, Pohang University of Science and Technology, Pohang 37673, Korea}

\date{\today}

\begin{abstract}
\texorpdfstring{$\mathrm{Fe}_{3}\mathrm{GeTe}_{2}$}{Lg} (FGT), a ferromagnetic van der Waals topological nodal line semimetal, has recently been studied. Using first-principles calculations and symmetry analysis, we investigate the effect of a uniaxial tensile strain on the nodal line and the resultant intrinsic anomalous Hall effect (AHE). Our results reveal their robustness to the in-plane strain. Moreover, the intrinsic AHE remains robust even for artificial adjustment of the atomic positions introduced to break the crystalline symmetries of FGT. When the spin-orbit coupling is absent, the nodal line degeneracy remains intact as long as the inversion symmetry or the two-fold screw symmetry is maintained, which reveal that the nodal line may emerge much more easily than previously predicted. This strong robustness is surprising and disagrees with the previous experimental report [Y. Wang \textit{et al.}, \color{blue}\href{https://onlinelibrary.wiley.com/doi/full/10.1002/adma.202004533}{Adv. Mater. \textbf{32}, 2004533 (2020)}\color{black}], which reports that a uniaxial strain of less than 1 \% of the in-plane lattice constant can double the anomalous Hall resistance. This discrepancy implies that the present understanding of the AHE in FGT is incomplete. The possible origins of this discrepancy are discussed.
\end{abstract}

\maketitle


\section{introduction}

Two-dimensional magnetic van der Waals (vdW) materials have been investigated intensely in recent years~\cite{KuoSciRep2016,LeeNanoLett2016,10,1,2,3,4,5,6,7,8,9,11,12,13,14,21,54}. Especially, \texorpdfstring{$\mathrm{Fe}_{3}\mathrm{GeTe}_{2}$}{Lg} (FGT) has attracted particular attention as a ferromagnetic topological nodal line semimetal candidate~\cite{14,20,55,56,57}. The coexistence of the ferromagnetic ordering and the nontrivial electronic topology makes interesting Berry phase phenomena within this material, such as the intrinsic anomalous Nernst effect~\cite{20,52} and the intrinsic anomalous Hall effect (AHE)~\cite{14,51,30}. This topological nodal line originates from the symmetries of the layered structure of FGT, which connect the {\it d}-orbitals of Fe atoms in the adjacent layers. As the nodal line degeneracy is orbital-driven, it appears in the vanishing spin-orbit coupling (SOC) limit and is tunable depending on the magnetization direction. The SOC-induced band gap becomes the largest when the spin orientation is completely out-of-plane. The most substantial Berry curvature appears then, resulting in a tremendous intrinsic AHE~\cite{14}.

Strain engineering, an efficient method for controlling the electronic structure of vdW materials, is actively being studied in many scientific branches from emerging quantum phenomena to next-generation information device technologies~\cite{40,41,42,43,53}. Studies have already dealt with changes in the magnetic properties~\cite{31,29,32,16,31} and transport properties~\cite{24,17,58,59} of FGT by strain. In particular, the experimental finding that a uniaxial strain of less than 1 \% can cause a twofold increase in the size of the AHE is particularly intriguing~\cite{29}. As the vast AHE within FGT comes from the band topology, this significant change seems to result from the symmetry breaking that occurred by strains. Also, manipulating the AHE through strain has already been explored in other two-dimensional magnetic materials such as \texorpdfstring{$\mathrm{Cr}\mathrm{I}_{3}$}{Lg} \cite{69,70} and \texorpdfstring{$\mathrm{Cr}\mathrm{Te}_{2}$}{Lg} \cite{71}. Therefore, it is interesting to investigate the strain-induced phenomena in FGT and eventually understand the principles underlying the AHE difference between FGT and other vdW mangets.


In this paper, we compute the intrinsic anomalous Hall conductivity (IAHC) in strained structures via the first-principles method to comprehend the impact of a uniaxial strain on the AHE within FGT. We also investigate how the nodal line is affected by strains. Surprisingly, our results show that uniaxial tensile strains do not significantly affect the intrinsic AHE and the nodal line degeneracy. Even when more constraints on the two bands forming the nodal line are broken by artificially modifying the atomic positions, the overall IAHC is not varied much. Also, when the SOC is excluded, so the time-reversal symmetry ($T$) exists, the nodal line is maintained unless either an inversion ($P$) or a two-fold screw axis symmetry ($\tilde{C}_{2z}=\left\{ C_{2z}|\frac{1}{2}\hat{z} \right\}$) is broken. This imposes a less stringent constraint on the symmetry as compared to the earlier theoretical study of the topological nodal line in FGT~\cite{14}. While the robustness of the intrinsic AHE and the nodal line is surprising in itself, it disagrees with the experimental result~\cite{29} that reports two-fold increase of the anomalous Hall resistance upon the strain of less than 1 \%. This discrepancy implies that our understanding of the AHE in FGT needs to be improved. The possible origins of the discrepancy are discussed.

The organization of this paper is as follows. In Sec.~\ref{sec2}, we introduce the uniaxial strains, artificial lattice distortions, the resultant changes in the symmetries, and the computational details. In Sec.~\ref{sec3}, we show how uniaxial strains change the IAHC and the nodal line, and discuss with a symmetry analysis. In Sec.~\ref{sec4}, we summarize our conclusions.


\section{\label{sec2} method}
\begin{figure}[t]
\includegraphics[width=3.375in]{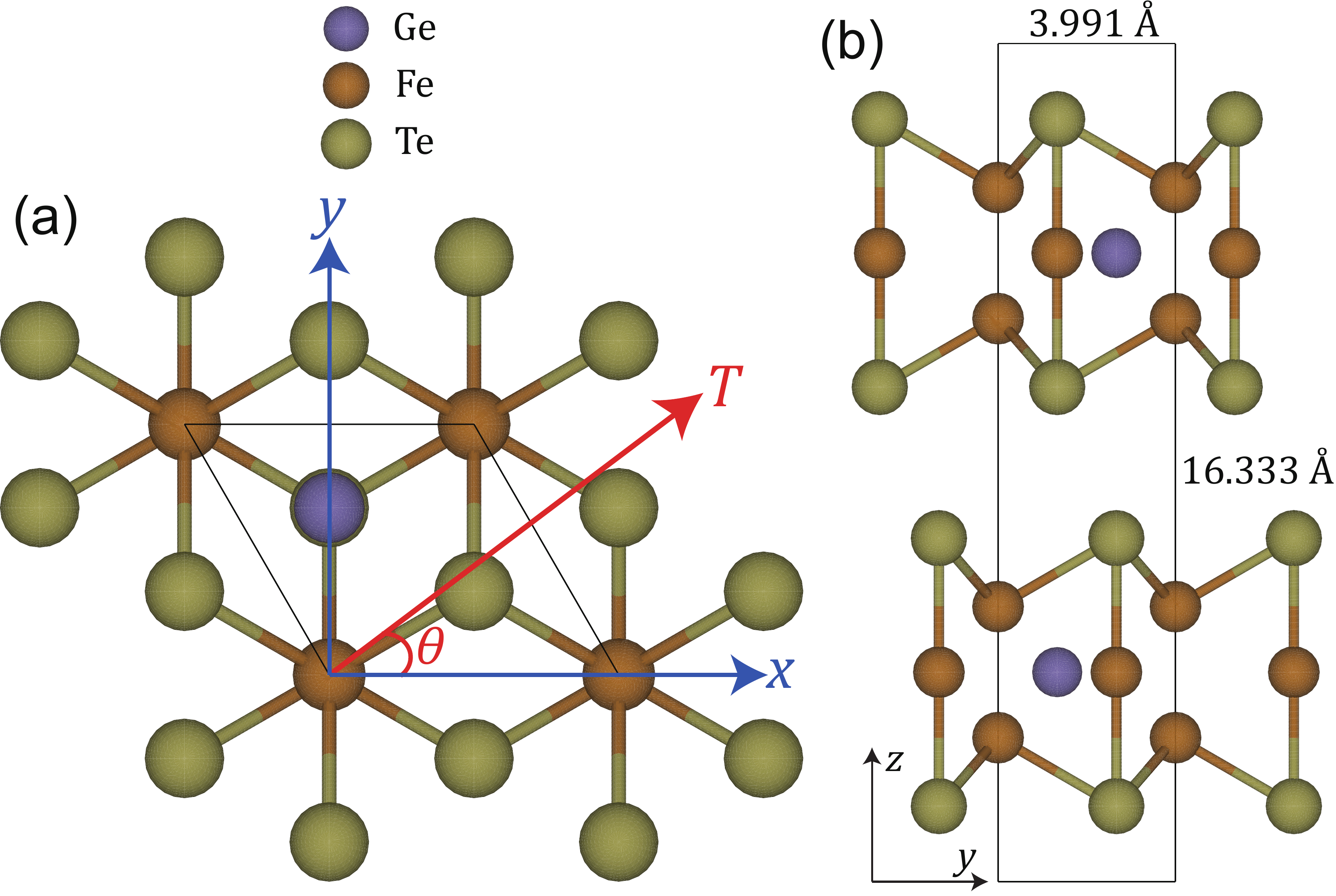}
\caption{\label{fig:fig1} Structure of the pristine FGT (a) viewed on the z-axis and (b) the x-axis. A uniaxial strain $T$ is indicated by an angle ($\theta$) to the x-axis. Considered strains are armchair (AC, $\pi/2$), zigzag (ZZ, 0), and chiral (CH, $\pi/12$).}
\end{figure}

Bulk crystalline FGT consists of AB-stacked alternating atomic layers of honeycomb lattices, where two Fe atoms are positioned vertically above and below the center of each hexagon [Fig.~\ref{fig:fig1}]. It belongs to the space group $D_{6h}^4$ ($P6_{3}/mmc$, No.194), whose generators are the six-fold screw rotation ($\tilde{C}_{6z}={C_{6z}|\frac{1}{2}\hat{z}}$), the two-fold rotation ($C_{2x}$), and the inversion ($P$). According to the previous study, the following combinations of the symmetries protect the two-fold nodal line degeneracy along the KH symmetry line: $\tilde{C}_{6z}\cdot P$, $M_{x}\cdot P$, and $P\cdot T$ (at the K point), $C_{3z}\equiv(\tilde{C}_{6z})^2$ and either $\tilde{C}_{6z}\cdot M_{x}$ or $P\cdot T$ (at any point between the K and H points), and $\tilde{M}_{z}\cdot P\cdot T$ (on the $k_{z}=\pi$ plane, including the H point)~\cite{14}. 

We consider tensile strains along three different directions ($\theta$) to break some of them: armchair (AC, $\pi/2$), zigzag (ZZ, 0), and chiral (CH, $\pi/12$) [Fig.~\ref{fig:fig1}(a)]. Since the unit cell is hexagonal, the AC strain along $\pi$/2 direction is equivalent to the strain along $\pi$/6 direction. The CH strain along $\pi$/12 direction is heading towards the center of the other two strains. The magnitude of the strains is between 0 $\sim$ 5 \% of the in-plane lattice constant. The strained FGT has lower symmetries than the pristine FGT. Table I shows the symmetries of each strained FGT.

Moreover, to examine the effect of the symmetry breaking further, we consider additional lattice distortions to 1 \% and 5 \% AC-strained structures to make them possess still lower symmetries. Because the two wave functions constituting the nodal line are mainly composed of {\it d}-orbitals of Fe, positions of Fe constituting the hexagonal lattices are chosen to be moved. Through this process, we make four different structures with the following crystalline symmetries: one with only $P$, one with only $\tilde{M}_{z}\cdot P$, one with only $\tilde{M}_{z}$, and one with all three of them. The size of movements is set equal to the extent to which Fe atoms move purely by the strains: 0.0173~\AA\ for the 1 \% strain and 0.0864~\AA\ for the 5 \% strain. We note that the lattice constants are adopted from the previous report~\cite{60}, and the atomic configurations are given in the supplemental material~\cite{supp} as Tables~SI-SIII.


\begin{table}[t]
\caption{\label{tab:symms} 
Existent symmetries in the AC-, ZZ-, and CH-strained FGT. $\tilde{M}_{i} \cdot P$ is equivalent to the two-fold screw axis symmetry ($\tilde{C}_{2i}$) for $i=x,y,z$.}
\centering
\begin{ruledtabular}
\begin{tabular}{ccccccccc}
AC/ZZ & $E$ & $P$ & $\tilde{M}_{z}$ & $\tilde{M}_{z}\cdot P$ & $M_x$ & $M_{x}\cdot P$ & $\tilde{M}_{y}$ & $\tilde{M}_{y}\cdot P$\\ 
CH    & $E$ & $P$ & $\tilde{M}_{z}$ & $\tilde{M}_{z}\cdot P$ \\ 
\end{tabular}
\end{ruledtabular}
\end{table}

First-principles calculations are composed of three steps. First, all structures are relaxed with total energy convergence threshold 1.36$\times 10^{-3}$ eV and force convergence threshold 0.0257 eV/\AA, while lattice parameters are fixed. Then the electronic structure of each relaxed structure is obtained. This step is performed by using QUANTUM-ESPRESSO \cite{44} package with the projector augmented wave pseudopotentials \cite{45} from PSlibrary \cite{46} and the revised Perdew-Burke-Ernzerhof exchange-correlation functional \cite{47}. The SOC is also considered. A Monkhorst-Pack \cite{48} $\textbf{k}$-grid of 20$\times$20$\times$5 is used, and the cutoff energy of wavefunctions is chosen to be 1225 eV. We set the magnetization direction to be the z-axis direction in Fig.~\ref{fig:fig1}.

Second, the maximally localized Wannier functions (MLWFs) are obtained from the Kohn-Sham states using the code WANNIER90 \cite{49}. We set the initial projections to be \textit{d$_{z^2}$}, \textit{d$_{xz}$}, \textit{d$_{yz}$}, \textit{d$_{x^2-y^2}$}, \textit{d$_{xy}$} for Fe, and \textit{p$_{z}$}, \textit{p$_{x}$}, \textit{p$_{y}$} for Ge and Te. From 178 Kohn-Sham states, 96 MLWFs are obtained. We set the frozen window as 2 eV above the Fermi energy for the disentanglement of inner and outer spaces. 

Third, the energy eigenvalue, the total Berry curvature ($\Omega_{\alpha\beta}$), and the IAHC ($\sigma_{\alpha\beta}^{\mathrm{AH}}$) are evaluated with the MLWFs. We use Kubo formula within the linear response theory to compute the total Berry curvature and the IAHC:
\begin{align}
    \sigma_{\alpha\beta}^{\mathrm{AH}} 
    &= \frac{e^{2}}{\hbar} \frac{1}{V_{cell} N_{k}} 
    \sum_{\textbf{k}} (-1) \Omega_{\alpha\beta}(\textbf{k}), \\
    \Omega_{\alpha\beta}(\textbf{k}) 
    &= \sum_{n} f_{n}(\textbf{k}) \Omega_{n,\alpha\beta}(\textbf{k}), \\
    \Omega_{n,\alpha\beta}(\textbf{k})
    &= -2 \mathrm{Im} \sum_{m \neq n} 
    \frac{v_{nm,\alpha}(\textbf{k})v_{mn,\beta}(\textbf{k})}
         {(\varepsilon_{n,\textbf{k}} - \varepsilon_{m,\textbf{k}})^2 + \Gamma^2}
\end{align}
where $v_{nm,\alpha}(\textbf{k})=\mel{n\textbf{k}}{\partial_{k_{\alpha}}\hat{H}(\textbf{k})}{m\textbf{k}}$ are the elements of the velocity operator, and $\Gamma$ is a smearing parameter whose unit is energy. We set $\Gamma$ and $k_{B} T$ in the Fermi-Dirac distribution function to be 0.0129 eV, corresponding to 150 K and less than the Curie temperature of bulk FGT~\cite{72,73}. The summation is performed over a uniform \textbf{k}-grid of 120$\times$120$\times$60.

To visualize the nodal line, we investigate the energy gap between the two bands forming the nodal line in the absence of the SOC. We compute on $k_{y}=0$ and several $k_{z}=l\pi\ (0\leq l\leq 1)$ planes near the KH symmetry line [Figs.~\ref{fig:fig3}(a) and \ref{fig:fig3}(c)].

\section{\label{sec3} Results}
\begin{figure}[!t]
\includegraphics[width=3.375in]{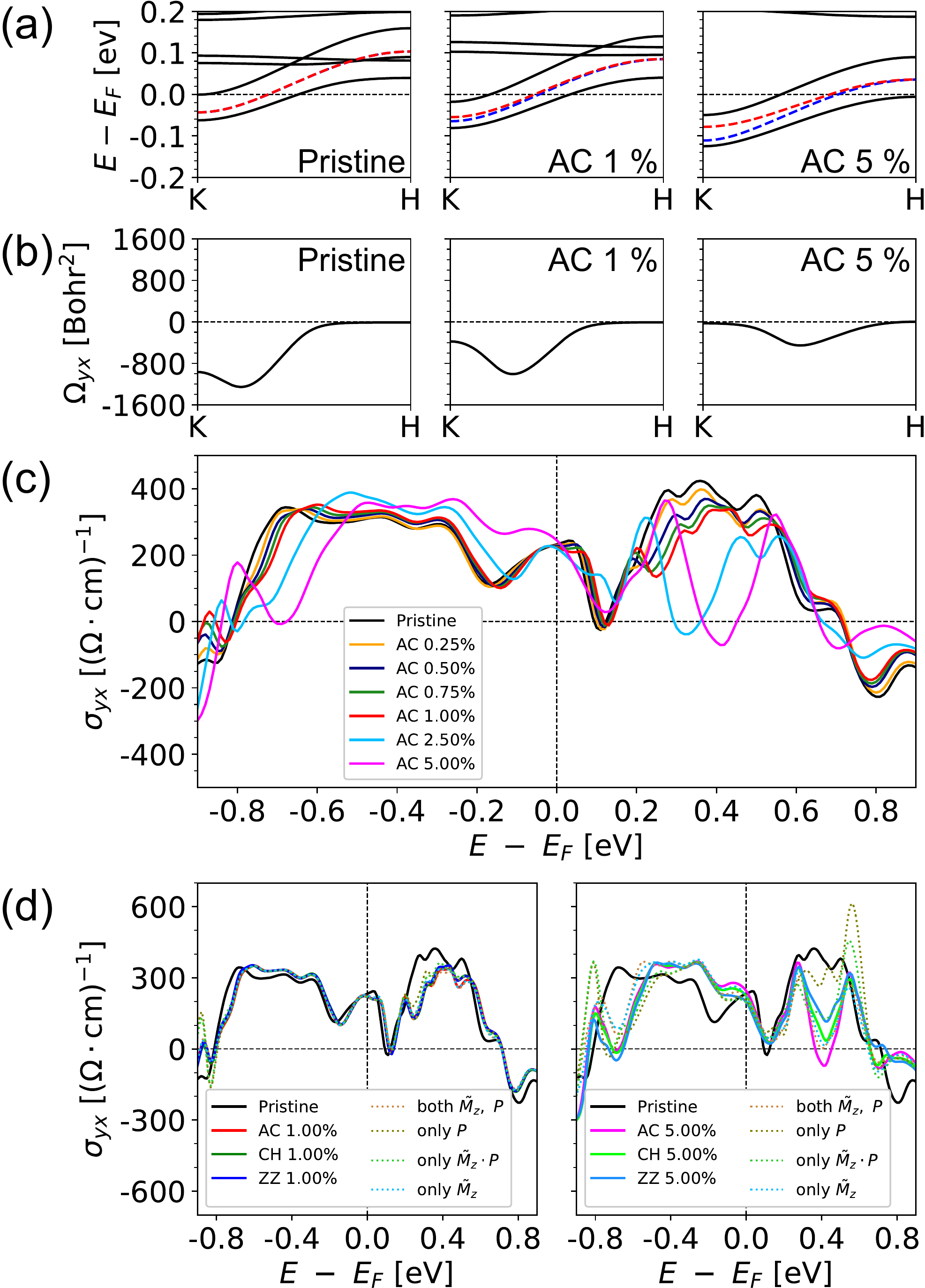}
\caption{\label{fig:fig2}(a) Band structures of the undeformed, 1 \%, and 5 \% AC-strained FGT along the KH symmetry line with SOC (solid, black) and without SOC (colored, dashed). (b) Corresponding total Berry curvatures. (c) The Fermi energy dependence of the intrinsic anomalous Hall conductivity (IAHC) in $x$ \% $ (0\leq x\leq 5)$ AC-strained FGT. (d) The Fermi energy dependence of the IAHC in 1 \% (left) and 5 \% (right) AC-, ZZ-, and CH-strained FGT. Dotted lines represent the IAHC obtained from the AC-strained FGT with additional lattice distortions that further lower the system's symmetry. In (c), results for different strain directions closely overlap and are almost indistinguishable.}
\end{figure}

Figure~\ref{fig:fig2}(a) shows the band structures of the pristine, 1 \%, and 5 \% AC-strained FGT. The red- and blue-colored bands are related to the two-fold nodal line degeneracy. As the magnitude of applied tensile strain increases, the gap between the two bands near the K point grows while it stays zero at the H point. This result is due to the breaking of $\tilde{C}_{6z}$ by the AC strain, required for the existence of the degeneracy at the K point and any point between the K and H points. However, as $\tilde{M}_{z}\cdot P\cdot T$ is unaffected by the AC strain, the degeneracy at the H point remains. When the SOC is engaged, it separates the colored bands so the red- and blue-colored bands become the black bands located just above and below them, respectively. These two bands contribute to the total Berry curvature with opposite signs. When the strain intensity increases, both bands shift towards below the Fermi energy. As a result, the total Berry curvature, a sum of the contributions of all bands below the Fermi energy, decreases as the magnitude of the strain increases [Fig.~\ref{fig:fig2}(b)].


In Fig.~\ref{fig:fig2}(c), we present the dependence of the IAHC on the Fermi energy, which is obtained from deformed structures by applying AC strains of various magnitudes. At the exact Fermi energy ($E-E_{F}=0$), the IAHC from the pristine FGT is 232.49 $(\Omega\cdot\textrm{cm})^{-1}$, a value consistent with previous experimental results [14, 17]. Notably, the IAHC remains almost constant at this point, irrespective of the intensity of the strain. In detail, as the strain strength increases to 2.5 \%, the IAHC decreases linearly by 5.3 \% to 220.23 $(\Omega\cdot\textrm{cm})^{-1}$. Conversely, for a higher strain of 5 \%, the IAHC increases to 244.71 $(\Omega\cdot\textrm{cm})^{-1}$, which is 5.3 \% larger than the case of the pristine FGT. Also, in a wide range of $E$, the IAHC is virtually unaffected by strains under 1 \%. Noticeable differences appear only for strong strains of 2.5 \% and 5 \%. To be precise, in the region just below the precise Fermi energy ($E-E_{F}\in [-0.2,0.0]$), where the intrinsic AHE is associated with the band topology and has already been experimentally implemented~\cite{18}, the 5 \% AC strain causes the IAHC to rise about 2.5 times higher than that obtained from the pristine FGT.

The dependence of the IAHC on the direction of the applied strain is illustrated in Fig.~\ref{fig:fig2}(d). As above, in the region where the intrinsic AHE of FGT is related to the topological nodal line, the IAHC obtained from 1 \% strained structures is nearly independent of the direction of the strain. Similarly, in the cases of 5 \% strains, although there is a contrast between the IAHC from them and the undeformed FGT, the directional difference among themselves in this region is insignificant. Also, even though the CH strain breaks more symmetries than the strains in the other two directions, the IAHC from the CH-strained FGT is between those from the AC- and ZZ-strained structures. These results may suggest that the existence of the nodal line degeneracy is not affected by any in-plane strain. The IAHC obtained from the lattice distorted structures shows a more exciting result [dotted lines in Fig.~\ref{fig:fig2}(d)]. Even though the symmetry constraints weaken, there is no apparent change in the IAHC. 

\begin{figure*}[!ht]
\includegraphics[width=\textwidth]{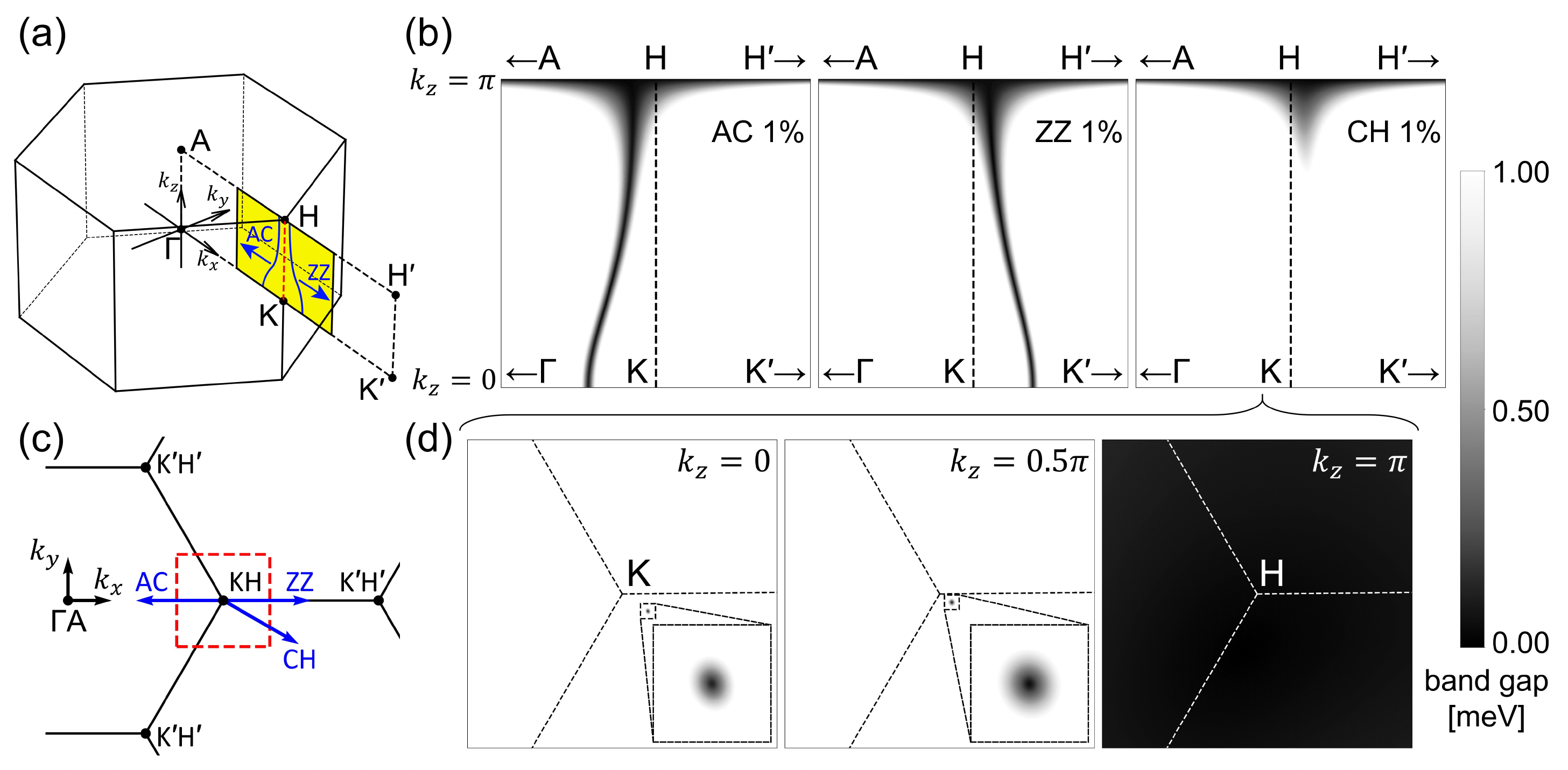}
\caption{\label{fig:fig3}
Effects of uniaxial strains on the nodal line degeneracy. (a) The Brillouin zone (BZ) of pristine FGT. The red-colored and dashed line represents the KH symmetry line. Two blue-colored lines schematically show the moved nodal lines due to the AC and ZZ strains. (b) The $\textbf{k}$-resolved band gap. The horizontal axis of each figure is a segment of a straight line connecting $\mathrm{\Gamma}$ and K, and width is the same as 3 \% of the distance between them. The Dashed line is equivalent to the KH symmetry line. (c) The BZ of pristine FGT viewed on the $k_{z}$ axis. Three blue-colored arrows show how a nodal point near the KH symmetry line moves on a $k_{z}$ plane when a uniaxial strain is engaged. (d) The $\textbf{k}$-resolved band gap on three $k_{z}$ planes in the case of 1 \% CH-strained FGT. The length of the axes is equal to 6 \% of the distance between the $\mathrm{\Gamma}$ and K points.}
\end{figure*}

In order to examine the effect of a strain on the nodal line, the energy difference between the two bands, responsible for the nodal line, is investigated in the absence of SOC. Before performing the calculations, we search for the region where the nodal line will be located in the momentum space through a symmetry analysis, which is given in the supplemental material~\cite{supp}. We find that in the presence of $\tilde{M}_{y}$, the degeneracy should appear on the $k_{y}=0$ plane [A$\Gamma$K$'$H$'$ plane, Fig.~\ref{fig:fig3}(a)]. To account for cases where $\tilde{M}_{y}$ does not exist such as the CH-strained FGT, we also investigate several $k_{z}$ planes [Fig.~\ref{fig:fig3}(c)].


Figure~\ref{fig:fig3}(b) shows how the nodal line moves on the $k_{y}=0$ plane due to the 1 \% AC, ZZ, and CH strains. In all cases, the degeneracy at $k_{z}=\pi$ plane is clearly preserved, since the Kramers' degeneracy in the plane is protected by the symmetry $\tilde{M}_{z}\cdot P$, which is not broken by any in-plane strain. As predicted by the symmetry analysis, the nodal line persists on the $k_{y}=0$ plane for the AC and ZZ strains, but disappears for the CH strain. Nevertheless, as there is a nodal point near the KH symmetry line on each $k_{z}\in [0,\pi]$ plane, their continuous connection can form the nodal line degeneracy [Figs.~\ref{fig:fig3}(c) and (d)]. For reference, the same results are also obtained in the case of 5 \% strain, given in the supplemental material~\cite{supp}. The results so far align with the previous report, which shows that nodal line is protected by $P$ and ($T$) in the absence of SOC~\cite{50}. In our situation, none of the considered strains disrupts $P$, and $T$ remains intact due to the exclusion of SOC. 

To go one step further, we also compute the band gap in lattice distorted structures and the detailed calculation results are given in the supplemental material~\cite{supp}. Here, we summarize the results. In the structure where both $P$ and $\tilde{M}_{z}$ exist (naturally their product $\tilde{M}_{z}\cdot P$ also exist), the nodal line appears just as in the CH-strained FGT. Although it is not located on the $k_{y}=0$ plane, nodal points appear on every $k_{z}$ plane. In the structure with only $P$, the $k_{z}=\pi$ plane is not any more a nodal plane due to the breaking of $\tilde{M}_{z}\cdot P$. However, a nodal point still appears on each $k_{z}$ plane. In the structure with only $\tilde{M}_{z}$ symmetry, the nodal line degeneracy is lifted in general. However, we find that the nodal line degeneracy persists even after the $\tilde{M}_{z}$ symmetry is broken, provided the $\tilde{M}_{z}\cdot P$ symmetry is maintained. Also, the Kramers' degeneracy at the $k_{z}=\pi$ plane is also clearly visible. Therefore, this nodal line cannot be distinguished from that  in the structure where both $P$ and $\tilde{M}_{z}$ exist.

From now on, we present a symmetry analysis that demonstrates how the combination of the two symmetries $\tilde{M}_{z}\cdot P$ and $T$ protects the topological nodal line within FGT, an expansion of the previous study~\cite{50} that shows that the combination of $P$ and $T$ protects the topological nodal line. The Hamiltonian of the strained FGT is modeled as shown below,
\begin{equation}
	H(\vec{k})
	=
	h_0(\vec{k}) \cdot \vec{\sigma_0} + \vec{h}(\vec{k}) \cdot \vec{\sigma},
	\label{eq:H_orig}
\end{equation}
where $\vec{k}=(k_x,k_y,k_z)$ and $\vec{\sigma}$ is the Pauli matrix in the bases of nodal line-related states $\psi_A$ and $\psi_B$, and $h_0(\vec{k})$ and $\vec{h}(\vec{k})$ are real coefficients. In the absence of the strain, the nodal line results from the two bands (both are superpositions of $\psi_A$ and $\psi_B$) touching with each other.
The operator $T\tilde{M_z}P$ maps the wavevector $(k_x, k_y, k_z)$ to $(k_x, k_y, -k_z)$ due to the actions of $\tilde{M_z} \cdot P$ and $T$, which transform $(k_x, k_y, k_z)$ into $(-k_x, -k_y, k_z)$ and $(-k_x, -k_y, -k_z)$, respectively.
The Schr\"{o}dinger equation for $H(\vec{k})$ is given by
\begin{equation}
H(\vec{k}) u_n(\vec{k}) = E_n(\vec{k}) u_n(\vec{k}),
\end{equation}
where
$u_n(\vec{k}) = [\psi_A(\vec{k}), \psi_B(\vec{k})]^T$ is the state of the $n$-th band, and $E_n(\vec{k})$ is the corresponding energy. The state is a superposition of the state $\psi_A$ in the A layer and the state $\psi_B$ in the B layer in the AB-stacked FGT.
To analyze the constraints on $H(\vec{k})$ due to the symmetries $\tilde{M_z}P$ and $T$, we examine their effect on the Schr\"{o}dinger equation. We have
\begin{equation}
	T\tilde{M_z}P H(\vec{k}) u_n(\vec{k}) = T\tilde{M_z}P E_n(\vec{k}) u_n(\vec{k}).
\end{equation}
It follows that
\begin{equation}
	H(\vec{k}') T\tilde{M_z}P u_n(\vec{k}) = E_n(\vec{k}) T\tilde{M_z}P u_n(\vec{k}).
\end{equation}
where $\vec{k}' = (k_x, k_y, -k_z)$.
Since $T$ can be replaced with the complex conjugate operator $K$, 
the Schr\"{o}dinger equation resulting from the action of  $\tilde{M_z}P$ and $K$ is represented as follows, using $\tilde{M_z}P u_n(\vec{k})=[\psi_B(\vec{-k}'), \psi_A(\vec{-k}')]^T$.
\begin{equation}
	H(\vec{k}')
	\begin{pmatrix}
		\psi_B^*(\vec{k}') \\
		\psi_A^*(\vec{k}')
	\end{pmatrix}
	= E_n(\vec{k})
	\begin{pmatrix}
		\psi_B^*(\vec{k}') \\
		\psi_A^*(\vec{k}')
	\end{pmatrix},
\label{eq:H_TMzP}
\end{equation}
where $E_n(\vec{k}) = E_n(\vec{k}')$.
Since
\begin{equation}
	\begin{pmatrix}
		\psi_B^*(\vec{k}) \\
		\psi_A^*(\vec{k}) 
	\end{pmatrix}
	=
	K
	\begin{pmatrix}
		\psi_B(\vec{k})  \\
		\psi_A(\vec{k}) 
	\end{pmatrix},
\end{equation}
we apply $K$ to both sides of Eq.~\eqref{eq:H_TMzP} to utilize the property $K^2=1$. This results in
\begin{eqnarray}
	K H(\vec{k}') K
	\begin{pmatrix}
		\psi_B(\vec{k}') \\
		\psi_A(\vec{k}')
	\end{pmatrix}
	=
	E_n(\vec{k})
	K^2
	\begin{pmatrix}
		\psi_B(\vec{k}') \\
		\psi_A(\vec{k}')
	\end{pmatrix}.
\label{eq:H_TMzP_T}
\end{eqnarray}
Furthermore, since
\begin{equation}
	\begin{pmatrix}
		\psi_B(\vec{k}) \\
		\psi_A(\vec{k})
	\end{pmatrix}
	=
	\sigma_x
	\begin{pmatrix}
		\psi_A(\vec{k}) \\
		\psi_B(\vec{k})
	\end{pmatrix},
\end{equation}
we add $\sigma_x$ to both sides of the Eq.~\eqref{eq:H_TMzP_T} to take advantage of the property that $\sigma_x^2=I_2$, resulting in
\begin{eqnarray}
	\sigma_x K
	H(\vec{k}')
	K \sigma_x
	\begin{pmatrix}
		\psi_A(\vec{k}') \\
		\psi_B(\vec{k}')
	\end{pmatrix}
	=
	E_n(\vec{k})
	\sigma_x^2
	\begin{pmatrix}
		\psi_A(\vec{k}') \\
		\psi_B(\vec{k}')
	\end{pmatrix}.
\label{eq:H_TMzP_MzP}
\end{eqnarray}
The left side of Eq.~\eqref{eq:H_TMzP_MzP} is $H(\vec{k}') u_n(\vec{k}')$, leading to the following relation of the Hamiltonian in momentum space:
\begin{equation}
	\sigma_x K H(\vec{k}') K \sigma_x = H(\vec{k}').
\label{eq:H_const}
\end{equation}
Using the relation given by
\begin{equation}
	K 
	\begin{pmatrix}
		\sigma_x   \\
		\sigma_y   \\
		\sigma_z  
	\end{pmatrix}
	K
	=
	\begin{pmatrix}
		\sigma_x   \\
		-\sigma_y   \\
		\sigma_z  
	\end{pmatrix},
\end{equation}
the coefficients of the Hamiltonian in Eq.~\eqref{eq:H_const} transform into
\begin{equation}
	\sigma_x \left[ h_0(\vec{k}')\sigma_0 + h_x(\vec{k}')\sigma_x - h_y(\vec{k}')\sigma_y + h_z(\vec{k}')\sigma_z \right] \sigma_x = H(\vec{k}').
\end{equation}
Furthermore, by using the relation of
\begin{equation}
	\sigma_x  
	\begin{pmatrix}
		\sigma_x   \\
		\sigma_y   \\
		\sigma_z  
	\end{pmatrix}
	\sigma_x
	=
	\begin{pmatrix}
		\sigma_x   \\
		-\sigma_y   \\
		-\sigma_z  
	\end{pmatrix},
	\label{eq:fin_MzP}
\end{equation}
we obtain
\begin{equation}
	h_0(\vec{k}')\sigma_0+h_x(\vec{k}')\sigma_x + h_y(\vec{k}')\sigma_y - h_z(\vec{k}')\sigma_z = H(\vec{k}').
\end{equation}
The following constraints on the coefficients of $H(\vec{k}')$ are derived:
\begin{eqnarray}
	h_{x}(\vec{k}')
	=
	h_{x}(\vec{k}'),
	\\
	h_{y}(\vec{k}')
	=
	h_{y}(\vec{k}'),
	\\
	h_{z}(\vec{k}')
	=
	-h_{z}(\vec{k}').
\end{eqnarray}
As a result, we derive $h_z(\vec{k})=0$, and the Hamiltonian and energy eigenvalue can be represented as
\begin{eqnarray}
	H(\vec{k})
	&=&
	h_0(\vec{k})  \vec{\sigma_0} + h_x(\vec{k}) \vec{\sigma_x} + h_y(\vec{k}) \vec{\sigma_y},
	\\
	E(\vec{k})
	&=&
	h_0(\vec{k})\pm\sqrt{h_x(\vec{k})^2+h_y(\vec{k})^2}.
\end{eqnarray}
Consequently, two energy eigenvalues become degenerate for suitable values of $k_x$ and $k_y$ that satisfy the two constraints $h_x(\vec{k})=0$ and $h_y(\vec{k})=0$ for a given value of $k_z$. This situation amounts to the band touching within the given $k_z$ plane. Since the number of free variables ($k_x$ and $k_y$) match the number of constraints, the band touching occurs generically for each $k_z$. By connecting the band touching points ($k_x$, $k_y$) for each $k_z$ from $k_z=0$ to $\pi$, one obtains the nodal line.

In the presence of uniaxial strains, the coexistence of $\tilde{M_z}\cdot P$ and $P$ leads to the formation of nodal lines. The AHE is related to the degeneracy of these nodal lines, which is lifted by SOC. Therefore, we expect the magnitude of the IAHC to remain largely unchanged in the presence of such strains. This behavior is observed when either $\tilde{M_z}\cdot P$ or $P$ are present. However, when $\tilde{M_z}$ is present alone, the nodal line degeneracy is lifted weakly at 1\% strain and significantly at 5\%, as shown in Figs. S17 and S19 in the supplementary material~\cite{supp}. Nevertheless, the IAHC remains largely unaffected at 1\% strain since the energy gap after the degeneracy lifting is small compared to the gap widened by SOC. 

Lastly, we compare our calculation results with the recent experimental result~\cite{29}. Contrary to our theoretical results, the experiment reports that the anomalous Hall resistance increases two-fold due to a weak strain of less than 1 \%. Thus, our calculation results disagree with the experimental result. Although the origin of this discrepancy is unclear, it is evident that the strain effect on the AHE in FGT requires further study both theoretically and experimentally. Here we discuss possible origins of the discrepancy. A possible origin is the AHE of extrinsic origin. While our calculation is limited to the AHE of intrinsic origin, the extrinsic contributions to the AHE may be important.  In transition-metal ferromagnets such as Fe, Co, and Ni, it has been demonstrated that the AHE can receive an extrinsic contribution of up to 10-30\% or more~\cite{WeischenbergPRL2011,TungPRB2012}. Therefore, it is plausible that a significant extrinsic effect may also be present in FGT, contrary to the commonly adopted assumption that the AHE in FGT is dominated by the intrinsic contribution.  Another possible origin is the interplay between the strain and the electron-electron interaction. The interaction effect may be considered in the density functional theory calculation by introducing the Coulomb interaction parameter $U$. For the pristine FGT, the IAHC does not change significantly regardless of whether $U$ is considered. For this reason, we excluded the contribution of $U$ in our density functional theory calculations. But, a more systematic study of the $U$ effect is necessary. Still another possible origin is the concurrent variation of the longitudinal resistivity $\rho_{xx}$ with the strain in the experiment~\cite{29}. Considering the relation between the anomalous Hall conductivity $\sigma_{yx}$ and the anomalous Hall resistivity $\rho_{yx}$,  $\sigma_{yx}=\rho_{yx}/(\rho_{xx}^2+\rho_{yx}^2)$, the discrepancy between our theoretical calculation results and the experimental result can be resolved if $\rho_{xx}$ increases concurrently with strain applied in the experiment since the simultaneous variations of $\rho_{yx}$ in the numerator and $\rho_{xx}^2+\rho_{yx}^2\approx \rho_{xx}^2$ in the denominator may cancel each other and leave $\sigma_{yx}$ unaltered. Further theoretical and experimental investigation is required to better understand the effect of the strain on the AHE in FGT. 

\section{\label{sec4} conclusions}
In this study, we have investigated the effect of a uniaxial strain on the intrinsic AHE and the topological nodal line within FGT first-principle calculations and model analysis. Our results demonstrate the robustness of both the AHE and nodal line against in-plane strain and artificial lattice distortions. Specifically, we have shown that nodal line degeneracy is preserved when either $P$ or $\tilde{M_z}\cdot P$ is conserved, indicating that nodal lines may be more easily observed than previously thought. Our findings suggest that the fundamental symmetries identified in this study can be utilized to replicate the topological AHE observed in FGT across a broad range of magnetic materials by selecting space groups that meet the symmetry criteria. Furthermore, our results reveal that the AHE is a robust property of the material, with a high degree of resistance to external perturbations such as strain. These findings have significant implications for the development and optimization of magnetic materials for a wide range of applications, including spintronics and quantum information.


\section*{acknowledgments}
\indent We thank Daegeun Jo, Seungyun Han and Wooil Yang for fruitful discussions. This research was supported by the National Research Foundation (NRF) of Korea (Grant No. 2020R1A2C2013484). Supercomputing resources including technical supports were provided by the Supercomputing Center, Korea Institute of Science and Technology Information (Contract No. KSC-2021-CRE-0283).

\bibliography{MAIN3}
\end{document}